\documentclass[aps,prd,nofootinbib,preprint,floatfix,unsortedaddress]{revtex4-1}
\usepackage{setspace}
\usepackage{graphicx}
\usepackage{dcolumn}
\usepackage{subfigure}
\usepackage{times,mathptm}
\usepackage{float}
\usepackage{color}
\usepackage{amsmath,amsfonts}
\usepackage{mathptmx}
\usepackage{mathrsfs}
\usepackage{bbm}
\usepackage{bm}
\usepackage{xfrac}

\newcommand{\bea}{\begin{eqnarray}}
\newcommand{\eea}{\end{eqnarray}}

\newcommand{\tJ}{\widetilde{J}}

\renewcommand{\d}{\delta}

\newcommand{\tf}{\widetilde{f}}
\newcommand{\tZ}{\widetilde{Z}}
\newcommand{\p}{\phi}
\renewcommand{\b}{\beta}

\newcommand{\tr}{\text{Tr}}
\newcommand{\hk}{\hat{k}}

\newcommand{\bx}{\mathbf{x}}
\newcommand{\by}{\mathbf{y}}

\newcommand{\m}{\mu}

\newcommand{\e}{\epsilon}

\renewcommand{\k}{\kappa}

\newcommand{\D}{\Delta}

\renewcommand{\th}{\theta}

\newcommand{\oh}{\frac{1}{2}}

\newcommand{\dg}{\dagger}
\newcommand{\non}{\nonumber}

\newcommand{\rf}[1]{(\ref{#1})}
\newcommand{\ra}{\rightarrow}
\newcommand{\pa}{\partial}

\usepackage{ulem}

\begin{document}

\bibliographystyle{h-physrev5}

\title{Mean field theory of effective spin models as a baryon fugacity expansion} 
 
\author{Jeff Greensite}
\affiliation{Niels Bohr International Academy, Blegdamsvej 17, DK-2100
Copenhagen \O, Denmark}
\altaffiliation[Permanent address: ]{Physics and Astronomy Dept., San Francisco State
University, San Francisco, CA~94132, USA}
\author{Kim Splittorff}
\affiliation{\singlespacing Discovery Center, Niels Bohr Institute, University of Copenhagen, Blegdamsvej 17, DK-2100
Copenhagen \O, Denmark}
\date{\today}
\vspace{60pt}
\begin{abstract}

\singlespacing
 
   The free energy of effective spin or ``Polyakov line" models with a chemical potential, based on the U($N$)
group, does not depend on the chemical potential.  In a mean field-inspired expansion, we show how the condition
of unit determinant, taking U($N$) to SU($N$), reintroduces the chemical potential, and allows us to express the
free energy, as a function of mean field variational parameters, in terms of an expansion in the baryon (rather than the quark) fugacity at each lattice site.  We solve the SU(3) mean field equations numerically to determine the phase diagram and compute observables.  We also calculate the first corrections to the leading order mean field results, and find that these can significantly shift the endpoint of a line of first order transitions.  The problem of deriving an effective spin model from full QCD is discussed.
\end{abstract}

\maketitle

\singlespacing
\section{\label{sec:intro}Introduction}

    Polyakov line or ``effective spin" models, with lattice actions of the form
\bea
S &=&   \b \sum_{x} \sum_{k=1}^d  [\tr U_x^\dg \tr U_{x+\hk} + \tr U_x  \tr U^\dg_{x+\hk}] 
  + \k \sum_x [e^\m \tr U_x + e^{-\m} \tr U^\dg_x] 
\label{action}
\eea
are of interest as crude models of gauge theories in $D=d+1$ dimensions at finite temperature and
chemical potential \cite{Karsch:1985cb}.  Indeed, actions of this form can be extracted from QCD directly by integrating out most of the variables via a combined strong-coupling and hopping parameter expansion, while keeping the Polyakov line holonomies $U_x$ fixed, and therefore \rf{action} is justified as an effective theory at least within the range of validity of these expansions.\footnote{Below we will refer to $\m$ in eq.\ \rf{action} as the ``quark" chemical potential, while keeping in mind the fact that, in the hopping parameter expansion, $\m$ is actually related to the quark chemical potential of full QCD by a factor of inverse temperature.}  At finite chemical potential $\mu$ the Polyakov line models have a sign problem, so that the usual Monte Carlo simulation is not directly applicable.  There are, nonetheless, several different methods which can be used to solve this model.  One of the earliest studies applied the complex Langevin equations to the SU(3) model
\cite{Karsch:1985cb,Bilic:1987fn,Aarts:2011zn}.  A second method is the mean field approach, applied to the
$\mu \ne 0$ case by Bilic et al.\ \cite{Bilic:1987fn}.  A third procedure, introduced in ref.\ \cite{Gattringer:2011gq}, is to convert the partition function to a ``flux" representation, which, in the SU($3$) case, has been simulated via the Prokof'ev-Svistunov worm algorithm by Mercado and Gattringer \cite{Mercado:2012ue}.  Finally, the model can also be solved, at least in some parameter range, by the reweighting technique \cite{Fromm:2011qi}.

    In this article we will revisit the mean field strategy, because there are certain aspects of that approach which we find illuminating.  It is generally believed that the free energy of effective spin models based on the U($N$) group do not depend on the chemical potential, for reasons discussed, e.g., in ref.\ \cite{Dumitru:2005ng}.  We first rederive this
result, in section \ref{UNP}, in the framework of a mean field-inspired expansion.  We then go on to show, in section \ref{SUN}, how the restriction to a unit determinant, which converts U($N$) to SU($N$), not only reintroduces the chemical potential, but also converts the mean field formulation into an expansion in baryon fugacity.  Numerical solutions of the mean field equations for the SU($3$) are presented in section \ref{SU3}, and the phase diagram (projected to the $\b-\m$ plane) is obtained.   We also display the effects of including the first correction to the mean field approximation. In section
\ref{Proposal} we present some comments on the problem of extracting the appropriate effective spin model from full QCD,  in the range of gauge couplings and quark masses of interest.  Our conclusions are in section \ref{Conclusions}.

\section{\label{UNP}U($N$) Polyakov line models}

    We will begin with models in which the effective spin (or ``Polyakov line") variable $U(x)$ is an element of the U($N$) group.  As already noted, the chemical potential disappears from the free energy in this case, but the example will set the stage for the more interesting SU($N$) models.

    Starting from the action \rf{action}, we mimic the mean field approach by first adding and subtracting constants $u,v$,
which will eventually become variational parameters:
\bea
S &=&    \b \sum_{x,k} \Bigl[(\tr U_x^\dg -v+v) (\tr U_{x+\hk}-u+u) 
 + (\tr U_x-u+u)  (\tr U^\dg_{x+\hk}-v+v)\Bigr] 
\non \\
& & \qquad + \k \sum_x [e^\m \tr U_x + e^{-\m} \tr U^\dg_x]
\non \\
&=& -2 \b dVuv + 2 \b dv \sum_x \tr U_x + 2 \b du \sum_x \tr U^\dg_x
 + \k \sum_x [e^\m \tr U_x + e^{-\m} \tr U^\dg_x] + J \ .
\eea
Here $V$ is the lattice volume, $d$ is its dimensionality, and
\bea
J &=&   \b \sum_{x,k} \Bigl\{ (\tr U_x^\dg -v)(\tr U_{x+\hk}-u) 
 + (\tr U_x -u)(\tr U^\dg_{x+\hk}-v) \Bigr\} \ .
\eea
We then have
\bea
S =  -2 \b dVuv + \sum_x [A_x \tr U_x + B_x \tr U^\dg_x] + J \ ,
\label{SJ}
\eea
where
\bea
A_x = A \equiv 2 \b dv + \k e^\m  ~~ \mbox{and}  ~~ B_x = B \equiv 2 \b du + \k e^{-\m} \ .
\eea
Although $A_x,B_x$ are $x$-independent constants, it is useful below to regard them as variables.  This allows us
to differentiate with respect to each of them, with the understanding that all the $A_x, B_x$ are set to $A$ and $B$, respectively, after the differentiation.

Ordinary mean field theory amounts to dropping $J$ in the action and, in the absence of a chemical potential,
setting, $u=v=m$, where $m$ is the mean field.  One then varies $m$ to minimize the free energy.   In our case,
define
\bea
Z_{mf} = e^{-F_{mf}}= e^{-2 \b dVuv} \prod_x \int dU_x \exp[A_x \tr U_x + B_x \tr U^\dg_x]
\label{Zmf0}
\eea
and
\bea
{Z\over Z_{mf}} = e^{-\Delta F} =  {\int DU  e^J \exp[\sum_x(A_x \tr U_x + B_x \tr U^\dg_x)] \over
\int DU  \exp[\sum_x(A_x \tr U_x + B_x \tr U^\dg_x)] }  \ .
\label{DelF}
\eea
Also defining the operator
\bea
\tJ\left[u,v,{\pa \over \pa A},{\pa \over \pa B}\right] &\equiv&
  \b \sum_{x,k} \Bigl\{ \left( {\pa \over \pa B_x} -v\right) \left( {\pa \over \pa A_{x+\hk}}-u\right) 
  + \left({\pa \over \pa A_x} -u\right)\left({\pa \over \pa B_{x+\hk}}-v\right) \Bigr\} \ ,
\eea
we have
\bea
\exp[-\Delta F] =
  \left( { e^{\tJ[u,v,{\pa \over \pa A},{\pa \over \pa B}]} \int DU  \exp[\sum_x(A_x \tr U_x + B_x \tr U^\dg_x)] \over \int DU  \exp[\sum_x(A_x \tr U_x + B_x \tr U^\dg_x)] } \right)_{|_{A_x=A,B_x=B}} \ .
\eea 

    Next we need to evaluate the U(N) integral
\bea
I &=& \int dU \exp[A \tr U + B \tr U^\dg] \ ,
\eea
which, by standard methods (cf.\ \cite{Kogut:1981ez}), becomes an angular integration
\bea
I &=& \int \prod_{n=1}^N {d \p_n \over 2\pi} {1\over N!} \e_{i_1...i_N} \e_{j_1...j_N}  e^{i(j_1-i_1)\p_1}...e^{i(j_N-i_N)\p_N}
  \exp\left[A \sum_{m=1}^N e^{i\p_m} + B\sum_{m=1}^N e^{-i\p_m}\right]
\non \\
&=& {1\over N!} \e_{i_1...i_N} \e_{j_1...j_N}   \prod_{n=1}^N \int {d \p_n \over 2\pi} e^{i(j_n-i_n)\p_n}
\exp\left[A e^{i\p_n} + B e^{-i\p_n}\right]
\non \\
&=&  {1\over N!} \e_{i_1...i_N} \e_{j_1...j_N}   \prod_{n=1}^N \left({\pa \over \pa A}\right)^{j_n}
\left({\pa \over \pa B}\right)^{i_n}
   \int {d \p_n \over 2\pi}\exp\left[A e^{i\p_n} + B e^{-i\p_n}\right]
\non \\
&=&  {1\over N!} \e_{i_1...i_N} \e_{j_1...j_N}  \prod_{n=1}^N \left({\pa \over \pa A}\right)^{j_n}
\left({\pa \over \pa B}\right)^{i_n} I_0\Bigl[2\sqrt{AB}\Bigr] \ .
\label{UN}
\eea
This gives us
\bea
Z_{mf} &=& e^{-2 \b dVuv} \prod_x  {1\over N!} \e_{i_1...i_N} \e_{j_1...j_N}  \prod_{n=1}^N 
\left({\pa \over \pa A_x}\right)^{j_n}
\left({\pa \over \pa B_x}\right)^{i_n} I_0\Bigl[2\sqrt{A_x B_x}\Bigr] \ .
\eea
   
   We now introduce rescaled variables
\bea
u &=& e^{-\m} u'  ~~\mbox{and}~~ v = e^\m v'
\non \\
A_x &=& (2 \b d v' +\k) e^\m = A'_x e^\m
\non\\
B_x &=& (2 \b d u' +\k) e^{-\m} = B'_x e^{-\m}
\non \\
{\pa \over \pa A_x} &=& e^{-\m} {\pa \over \pa A'_x}
\non \\
{\pa \over \pa B_x} &=& e^{\m} {\pa \over \pa B'_x} \ .
\label{rescale}
\eea
Then $Z_{mf}$ becomes
\bea
Z_{mf} &=& e^{-2 \b dVu' v'} \prod_x  {1\over N!} \e_{i_1...i_N} \e_{j_1...j_N}     \prod_{n=1}^N 
e^{(i_n-j_n)\m} \left({\pa \over \pa A'_x}\right)^{j_n}
\left({\pa \over \pa B'_x}\right)^{i_n} I_0\Bigl[2\sqrt{A'_x B'_x}\Bigr]
\non \\
&=& e^{-2 \b dVu' v'} \prod_x  {1\over N!} \e_{i_1...i_N} \e_{j_1...j_N}  
\exp\left[\left(\sum_{m=1}^N i_m - \sum_{m=1}^N j_m\right)\m \right]
\non \\
& & \qquad \times \prod_{n=1}^N 
\left({\pa \over \pa A'_x}\right)^{j_n}
\left({\pa \over \pa B'_x}\right)^{i_n} I_0\Bigl[2\sqrt{A'_x B'_x}\Bigr] \ .
\eea
At this point we note that, because of the $\e_{i_1...i_N} \e_{j_1...j_N}$  term,
\bea
\sum_{m=1}^N i_m = \sum_{m=1}^N j_m  \ .
\eea
Therefore 
\bea
Z_{mf} &=& e^{-2 \b dVu' v'} \prod_x  {1\over N!} \e_{i_1...i_N} \e_{j_1...j_N}     \prod_{n=1}^N 
 \left({\pa \over \pa A'_x}\right)^{j_n}
\left({\pa \over \pa B'_x}\right)^{i_n} I_0\Bigl[2\sqrt{A'_x B'_x}\Bigr]
\non \\
&=& e^{-2 \b dVu' v'} \prod_x \det\left[ \left({\pa \over \pa B'_x}\right)^i \left({\pa \over \pa A'_x}\right)^j 
 I_0\Bigl[2\sqrt{A'_x B'_x}\Bigr] \right] \ .
\eea
As a function of the rescaled variational parameters $u',v'$, $Z_{mf}$ is clearly, $\mu$-independent, and of course
it will remain $\m$-independent when $F_{mf}$ is minimized with respect to $u',v'$.  Likewise, all $\mu$ dependence cancels in the $\tJ$ operator
\bea
\tJ\left[u,v,{\pa \over \pa A},{\pa \over \pa B}\right] &=&     \b \sum_{x,k} \left\{ \left( {\pa \over \pa B_x} -v\right) \left( {\pa \over \pa A_{x+\hk}}-u\right) 
+ \left({\pa \over \pa A_x} -u\right)\left({\pa \over \pa B_{x+\hk}}-v\right) \right\}
\non \\
&=&    \b \sum_{x,k} \left\{ \left( {\pa \over \pa B'_x} -v'\right) \left( {\pa \over \pa A'_{x+\hk}}-u'\right) 
+ \left({\pa \over \pa A'_x} -u'\right)\left({\pa \over \pa B'_{x+\hk}}-v'\right) \right\} \ .
\non \\
\eea
From this we can conclude that both $F_{mf}$ and $\D F$, and therefore the free energy $F=F_{mf}+\D F$ itself,
are independent of the chemical potential $\mu$ in Polyakov line models based on the group U($N$).\footnote{A slight
subtlety is that at $\k=0$, the free energy depends not on $u',v'$ separately, but only on the product $u'v'=uv$.
Then one must appeal to the hermiticity of the action to set $u=v$.  For any non-zero $\k$ and $\m$, however, there is no such degeneracy.}

Before proceeding to SU($N$), we note that the expression for $Z_{mf}$ can be simplified a little further, using the
identity
\bea
             {\pa \over \pa A}  {\pa \over \pa B}  I_0\Bigl[2\sqrt{A B}\Bigr] =  I_0\Bigl[2\sqrt{A B}\Bigr]  \ ,
\eea
which is evident from the fact that
\bea
             I_0\Bigl[2\sqrt{A B}\Bigr] = \int {d\p \over 2 \pi} e^{A e^{i\p} + B e^{-i\p}} \ .
\eea
Then, defining the derivative operator
\bea
D_{ij}(x) = \left\{ \begin{array}{cl}
                    \left( {\pa \over \pa B'_x} \right)^{i-j} & i \ge j \cr
                     \left( {\pa \over \pa A'_x} \right)^{j-i} & i < j  \end{array} \right. \ ,
\eea
we may write
\bea
Z_{mf} = e^{-2 \b dVu' v'} \prod_x \det\left[ D_{ij}(x) I_0[2\sqrt{A'_x B'_x}] \right]
\eea
and
\bea
e^{-\D F} = \left( {1\over Z_{mf}} e^{\tJ[u',v',{\pa \over \pa A'},{\pa \over \pa B'}]} 
Z_{mf} \right)_{|_{A'_x=A',B'_x=B'}} \ .
\eea
Again, the $\m$-independence of the free energy is manifest.\footnote{This $\mu$-independence was also demonstrated in the $N=\infty$ limit in \cite{Christensen:2012km}.}

\section{\label{SUN}SU($N$) Polyakov line models}

    We can convert the U($N$) models considered above to SU($N$) models by simply converting the U($N$) group integration in eq.\ \rf{UN} to an SU($N$) integration. To accomplish this, we only have to insert a periodic delta 
function into the angular integrations, which imposes the constraint that $\sum_n \p_n = 0$ mod $2\pi$.  We use the identity
\bea
           \d_p\left(\sum_{n=1}^N \p_n\right) = {1\over 2\pi} \sum_{s=-\infty}^\infty \exp\left[ is \sum_{n=1}^N \p_n \right] \ .
\label{pdelta}
\eea
This introduces into each $\p_n$ integration an additional factor of $\exp[i s \p_n]$.  Tracing through the steps
of the previous section, we arrive at 
\bea
Z_{mf} =  e^{-2 \b dVu v} \prod_x  {1\over 2\pi} \sum_{s=-\infty}^\infty {1\over N!} \e_{i_1...i_N} \e_{j_1...j_N}     
\prod_{n=1}^N  \left({\pa \over \pa A_x}\right)^{j_n}
\left({\pa \over \pa B_x}\right)^{i_n} 
\left\{\begin{array}{cc}
(s \ge 0) &  \left({\pa \over \pa A_x}\right)^s \cr
(s < 0) &  \left({\pa \over \pa B_x}\right)^{|s|} \end{array} \right. 
 I_0\Bigl[2\sqrt{A_x B_x}\Bigr] \ .
\non \\
\eea
Now expressing everything in terms of the rescaled variables of eq.\ \rf{rescale}, this becomes
\bea
Z_{mf} &=&  e^{-2 \b dVu' v'} \prod_x  {1\over N!} \e_{i_1...i_N} \e_{j_1...j_N} {1\over 2\pi} \left\{ 
\sum_{s\ge 0} e^{-s N \m}  \prod_{n=1}^N 
 \left({\pa \over \pa A'_x}\right)^{s+j_n} \left({\pa \over \pa B'_x}\right)^{i_n} \right.
\non \\
& & \left. \qquad + \sum_{s < 0} e^{|s| N \m}  \prod_{n=1}^N 
 \left({\pa \over \pa A'_x}\right)^{j_n} \left({\pa \over \pa B'_x}\right)^{i_n + |s|} \right\}
  I_0\Bigl[2\sqrt{A'_x B'_x}\Bigr] \ .
\eea
Defining
\bea
D^s_{ij}(x) \equiv \left\{ \begin{array}{cc} 
      D_{i,j+s}(x) & s \ge 0 \cr
      D_{i+|s|,j}(x) & s<0 \end{array} \right. \ ,
\eea
we can express $Z_{mf}$ compactly in the form
\bea
Z_{mf} =  e^{-2 \b dVu' v'} (2\pi)^{-V} \prod_x  \sum_{s=-\infty}^\infty e^{s N \m} 
\det\Bigl[D^{-s}_{ij}I_0[2\sqrt{A'_x B'_x}]\Bigr]  \ ,
\label{Zmf}
\eea
where we have also changed variables $s\ra -s$ in the sum.  As before
\bea
e^{-\D F} = \left( {1\over Z_{mf}} e^{\tJ[u',v',{\pa \over \pa A'},{\pa \over \pa B'}]} 
Z_{mf} \right)_{|_{A'_x=A',B'_x=B'}} \ .
\eea
This gives a formal expression for the full free energy, $F(\m)=F_{mf}(\m)+\D F(\m)$ in terms of the variational
parameters $u',v'$, which should be chosen to minimize $F(\m)$.  

   The mean field expression for the free energy $F_{mf}$, as a function of the variational parameters $u',v'$ (or equivalently $A',B'$) has some features which are worth noting.  In the first place, the mean field partition function $Z_{mf}$ has now been expressed in terms of a product, at each site, of a fugacity expansion of the form
\bea
             \sum_{s=-\infty}^\infty e^{s N \m} 
\det\Bigl[D^{-s}_{ij}I_0[2\sqrt{A' B'}]\Bigr]  \ .
\eea
Here we see that the quark chemical potential $\m$ only occurs in the combination $N\mu$, which is, in effect,
the baryon chemical potential.  So in fact we have an expansion in the baryon, rather than quark, fugacity. 
In ref.\ \cite{Danzer:2012vw}, the determinant in an expansion of this sort is referred as
the ``canonical determinant." The second point is that parameter $s$, originally introduced in the representation 
\rf{pdelta} of the periodic delta function, has now emerged as the baryon number (which, if negative, is the number of antibaryons) per site.

   Of course, one still has to minimize the free energy with respect to the variational parameters, and this will introduce
some $N\mu$-dependence into the canonical determinants.  Strictly speaking, it is the mean field expression of the partition function as a function of (freely varying) parameters $u',v'$ which has the form of a fugacity expansion.  

   Successive improvements to the leading mean field result would be obtained by expanding the operator $\exp[\tJ]$ in a Taylor series.  In the case that $\k=0$, and $\b$ is so small that the minimum free energy is obtained at $u'=v'=0$, then the Taylor series simply generates the strong-coupling expansion.  At larger $\b$ and $\k$, the series also generates corrections to the leading mean-field result.  We will compute the effect of the leading correction in the next section.

    At this point, we should draw attention to the similarities and differences between our approach and the much earlier
work of Bilic et al.\  \cite{Bilic:1987fn}.  The starting point of the mean field treatment in \cite{Bilic:1987fn} was the action 
\rf{SJ} without the $J$-term.  The SU(3) group integral was expanded as a power series in $A,B$, and for this reason it was not obvious that the partition function is an expansion in baryon fugacity, arising from the unit determinant condition.   
In the next section we determine the phase diagram (for both real and imaginary $\m$), which was not displayed in 
\cite{Bilic:1987fn}, and work out leading corrections to the mean field result.

 \section{\label{SU3}Numerical results for the SU(3) Polyakov line model}
 
    We will now specialize to SU(3).  From eq.\ \rf{Zmf}, we see that the mean field free energy per lattice site at $N=3$ is
\bea
f_{mf} =  2 \b d u' v' - \log\left[  \sum_{s=-\infty}^\infty e^{3 s \m} 
\det\Bigl[D^{-s}_{ij}I_0[2\sqrt{A' B'}]\Bigr] \right] \ .
\eea
where we have dropped an irrelevant constant.
In numerical work we cannot sum $s$ over the full range $[-\infty,\infty]$, so it is necessary to cut off the sum
at some maximum baryon/antibaryon number $s_{max}$ per site
\bea
f_{mf} &\approx&  2 \b d u' v' - \log[G(A',B')] \ ,
\non \\
G(A',B') &\equiv&  \sum_{s=-s_{max}}^{s_{max}} e^{3 s \m} \det\Bigl[D^{-s}_{ij}I_0[2\sqrt{A' B'}]\Bigr] \ ,
\label{G}
\eea
and of course it is important, when computing observables, to check sensitivity to the cutoff.  We will return to 
this issue below.

   Minimizing the free energy with respect to the variational parameters $u',v'$, or, equivalently, with respect to
$A'=2 \b dv' + \k, B'=2 \b du' +\k$, leads to two equations
\bea
           {B'- \k \over 2 \b d} - {1 \over G(A',B')} {\pa G \over \pa A'} &=& 0 
\non \\           
           {A'-\k \over 2 \b d} - {1 \over G(A',B')} {\pa G \over \pa B'} &=& 0          
\label{roots}
\eea
whose roots may be determined numerically.\footnote{We have found Mathematica convenient for this purpose.}  At the minimum, we can regard $A'=A'(\b,\k,\m)$ and $B'=B'(\b,\k,\m)$ as functions of the parameters of the theory.  

   Apart from the free energy itself, the observables of interest are $\tr[U], \tr[U^\dg]$, and the baryon number
density $n$ (baryon number per lattice site).  The latter is given by  
\bea
             \langle n \rangle &=& - {\pa f_{mf} \over \pa (3\m)}
\non \\
&=& {1\over  G(A',B')}  \sum_{s=-s_{max}}^{s_{max}} s e^{3 s \m} \det\Bigl[D^{-s}_{ij}I_0[2\sqrt{A' B'}]\Bigr] 
\non \\
& & -{1\over 3}  \left( {\pa A'(\b,\k,\m) \over \pa \m} {\pa \over \pa A'} + {\pa B'(\b,\k,\m) \over \pa \m} {\pa \over \pa B'}\right) 
f_{mf}(A',B') \ ,
\eea
where it is understood that the derivative is taken at the point where $f_{mf}(A',B')$ is minimized.  But at this point,
the first derivatives of $f_{mf}$ with respect to $A'$ and $B'$ vanish.  Therefore
\bea
 \langle n \rangle =  {1\over  G(A',B')}  \sum_{s=-s_{max}}^{s_{max}} s e^{3 s \m} 
 \det\Bigl[D^{-s}_{ij}I_0[2\sqrt{A' B'}]\Bigr]   \  . 
 \eea
 
   From \rf{Zmf0} we see that
\bea
\langle \tr U \rangle &=& {1\over V}\sum_x {\pa \over \pa A_x} \log Z_{mf}
\non \\
&=&  {\pa \over \pa A} \log G(A',B')
\non \\
&=& e^{-\m} {\pa \over \pa A'} \log G(A',B') \ .
\eea
At the minimum of the free energy, determined by the roots of \rf{roots},  this simply becomes
\bea
\langle \tr U \rangle \equiv {1\over V} \sum_x \langle \tr U_x \rangle = e^{-\mu} u' = u \,
\eea
and likewise
\bea
\langle \tr U^\dg \rangle \equiv {1\over V} \sum_x \langle \tr U^\dg_x \rangle= e^{\mu} v' = v \ .
\eea
This is, of course, reminiscent of the standard mean field approach to a spin system, in which the variational parameter becomes the average spin.  It must be understood, however, that due to the complex weight there is no constraint that the ``average" values of $\tr U$ and $\tr U^\dg$ are necessarily bounded by $\tr  \mathbbm{1}$. 

\begin{figure}[ht]
\centering
\subfigure[~ Polyakov lines vs.\ $\b$.]{
\resizebox{70mm}{!}{\includegraphics{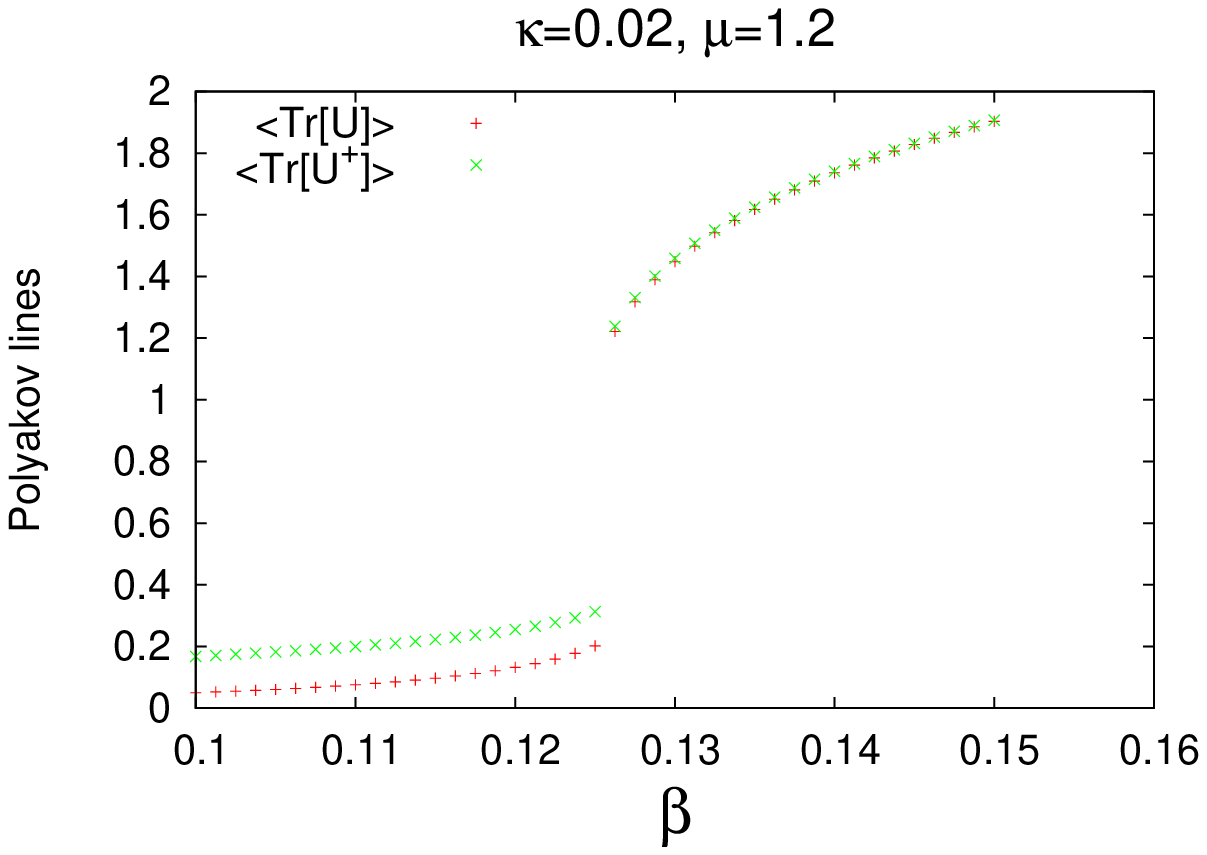}}
\label{uv_eps}
}
\subfigure[~ - free energy/site vs.\ $\b$.]{
\resizebox{70mm}{!}{\includegraphics{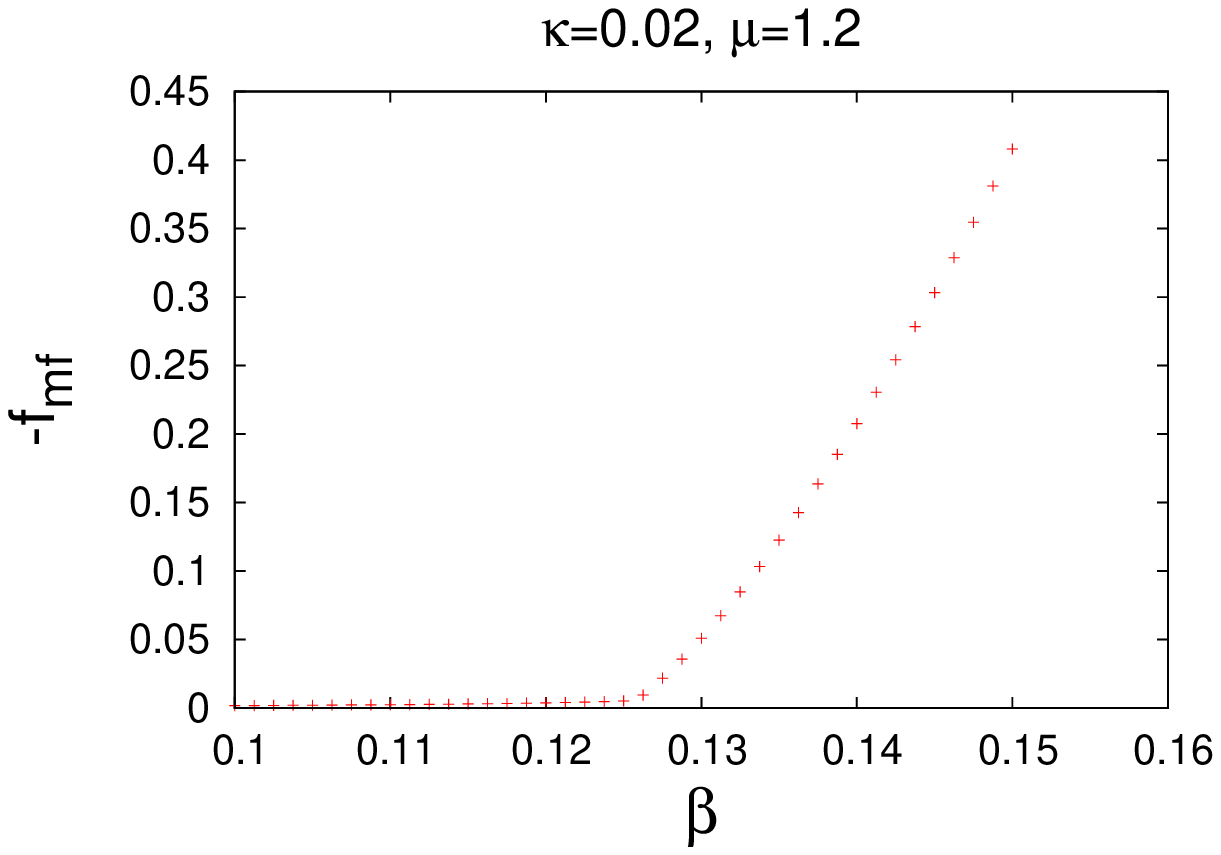}}
\label{free_eps}
}
\caption{Observables vs.\ $\b$ at fixed $\k=0.02$ and $\m=1.2$, evaluated at mean field level for SU(3).}
\label{vsb}
\end{figure}

   We now have all the tools needed to evaluate observables and map out the phase diagram.  Figure \ref{vsb}
shows a typical result for $\langle \tr U \rangle, \langle \tr U^\dg \rangle$ and the mean field free energy per site $f_{mf}$,
as a function of $\b$, at fixed $\k=0.02$ and chemical potential $\m=1.2$.  There is a clear first order phase transition
at $\b=0.1257$.  As the chemical potential is increased at fixed $\k=0.2$, the discontinuity at the transition decreases,
until it disappears altogether at $\mu=1.67$.  At larger $\mu$, there is only a crossover.
   
\begin{figure}[t!]
\centerline{\scalebox{1.0}{\includegraphics{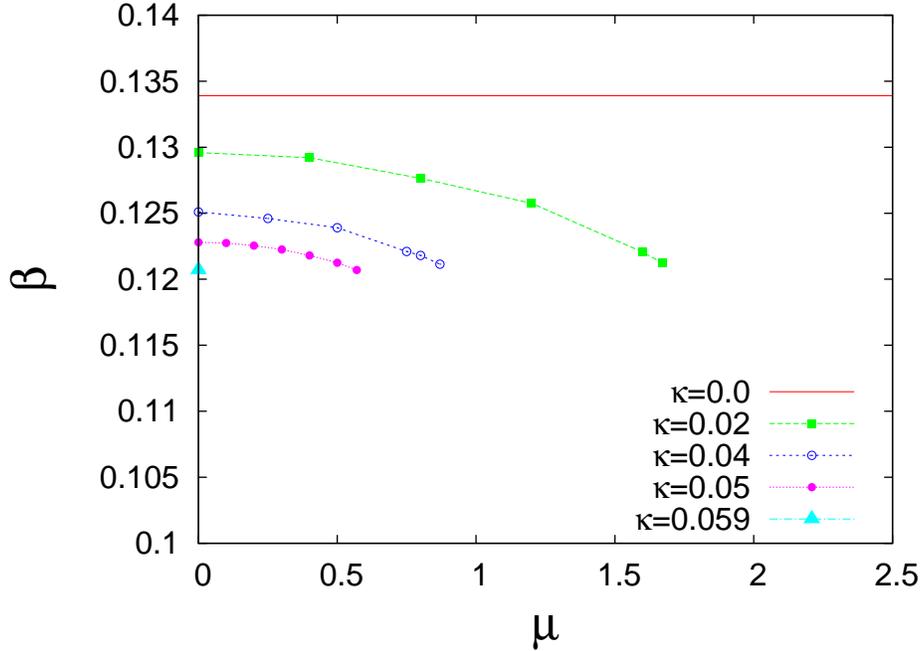}}}
\caption{Phase diagram of the Polyakov line model \rf{action} for the SU(3) group, obtained via mean field methods, in the $\b-\m$ plane at several values of $\k$.  The lines indicate first order transitions.  Beyond $\k=0.059$, there are no transitions at any value of $\mu$.}
\label{phase}
\end{figure}

   Repeating this procedure, we can map out the region of first order transitions in the $\b,\mu,\k$ parameter space.
In Fig.\ \ref{phase} we show sample first-order phase transition lines in the $\b - \m$ plane at $\k=0,0.02,0.04,0.05,0.059$.  At $\k=0$ the transition, at $\b=0.1339$, is of course independent of $\mu$.  At fixed, finite $\k$ the transition line terminates at some value of $\mu$, and this termination point happens at smaller and smaller values of $\m$ as $\k$ increases.  The transition line shrinks to a point at $\m=0$ for $\k=0.059$, and beyond this value of $\k$ there are no further transitions.

\begin{figure}[htb]
\centerline{\scalebox{1.0}{\includegraphics{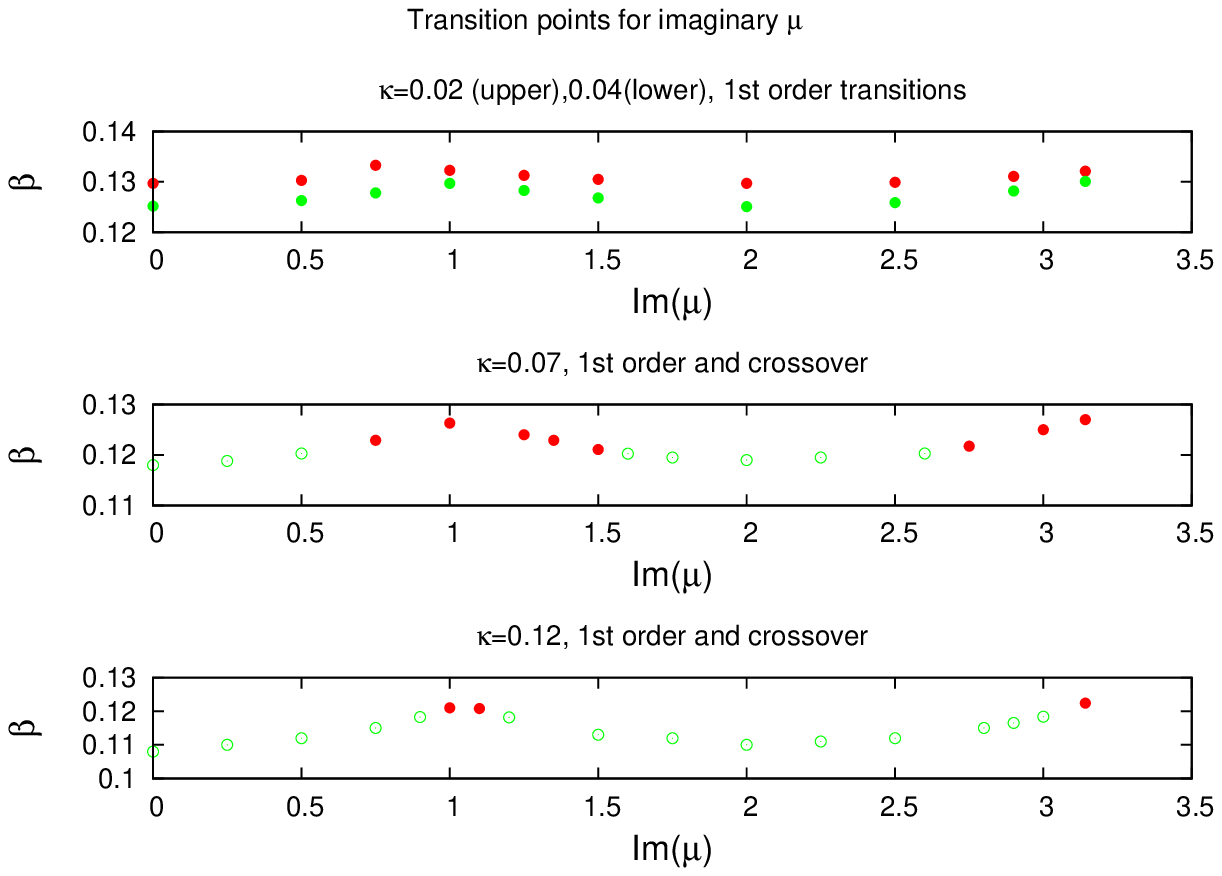}}}
\caption{Some transition points for Polyakov line models in the $\b-$Im($\m$) plane, for imaginary values of the chemical potential, at several values of $\k$.  Filled circles indicate first order transition points, open circles indicate a crossover.}   
\label{imphase}
\end{figure}

   We can also solve the mean field equations for imaginary $\m$.   The results for several values of $\k$ are shown
in Fig.\ \ref{imphase}.  The continuity of first order transition lines, as $\m$ varies from real to imaginary values, ties in with the considerations of ref.\ \cite{deForcrand:2010he}.

   Figure \ref{phase} can be compared directly to the phase diagram recently obtained by Mercado and Gattringer \cite{Mercado:2012ue}
via a Monte Carlo simulation in the flux representation.  The two diagrams are qualitatively, and even quantitatively, very similar. The main difference is that we only show first order transition points, and most of these are found in 
ref.\ \cite{Mercado:2012ue} to be
crossover points, rather than first order transitions.  According to Mercado and Gattringer \cite{Mercado:2012ue},
the endpoint of a line of first order transitions, at a given $\k$, occurs at a much smaller value of $\m$ than we find in our mean field calculations.  So an interesting question is whether inclusion of higher order corrections, beyond the leading order mean field result, would bring our endpoints to smaller values of $\mu$, in closer agreement with \cite{Mercado:2012ue}.  We will turn to this question in subsection \ref{corr} below.

\subsection{Effect of the baryon number cutoff}

   The data displayed above was obtained using a cutoff $s_{max}=4$ in the sum over baryon number, but the results
shown are quite insensitive to increasing the cutoff to $s_{max}=6$, and even to decreasing the limit to
$s_{max}=2$.   The reason for this insensitivity is that the phase transitions occur at values of the baryon number
density which are very small compared to the cutoff.  Only when the chemical potential is raised to values such
that the number density becomes comparable to $s_{max}$ does the cutoff dependence become apparent.  To illustrate
this dependence, we fix $\b=0.1257$ and $\k=0.02$ (where we have found a transition at $\mu=1.2$), and compute
the Polyakov lines and number density over a wider range of $\mu$.

\begin{figure}[ht]
\centering
\subfigure[~ $\langle \tr U \rangle$ vs.\ $\m$.]{
\resizebox{70mm}{!}{\includegraphics{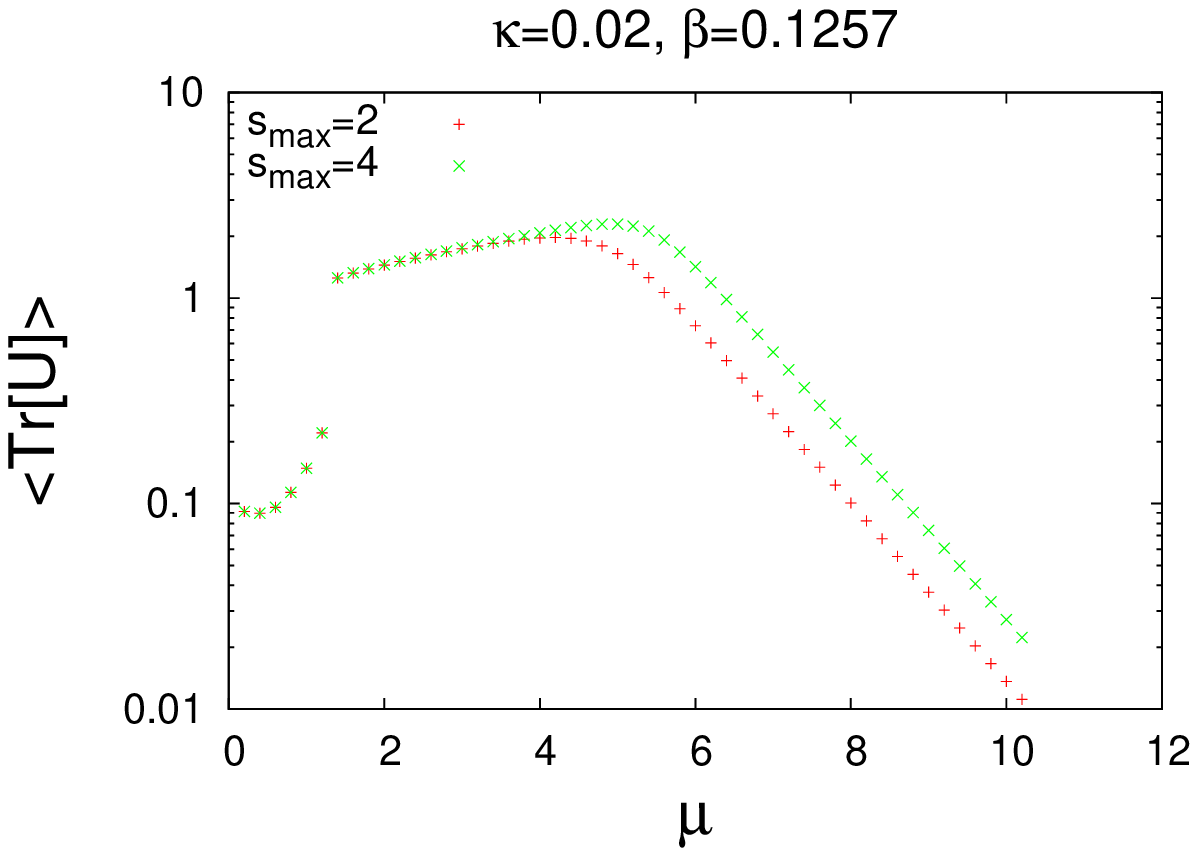}}
\label{u}
}
\subfigure[~  $\langle \tr U^\dg \rangle$ vs.\ $\m$.]{
\resizebox{70mm}{!}{\includegraphics{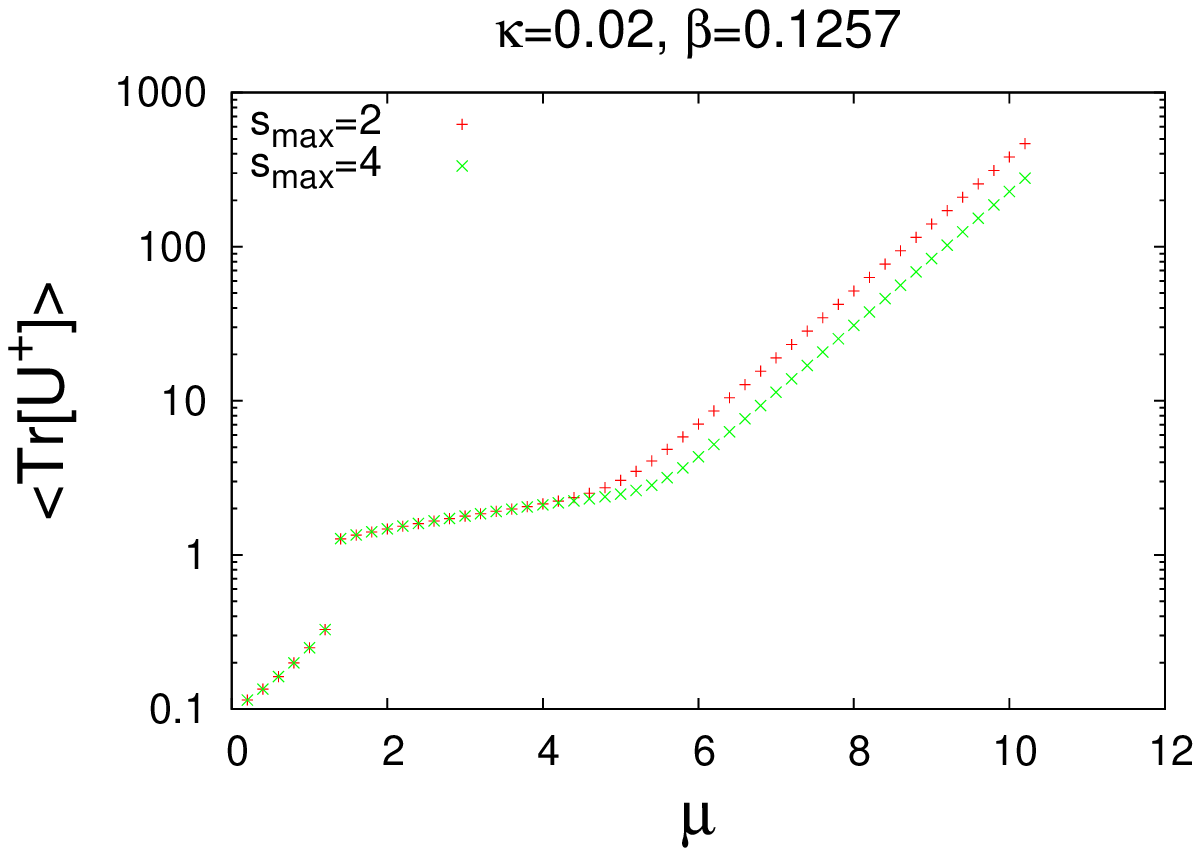}}
\label{v}
}
\subfigure[~  number density vs.\ $\m$.]{
\resizebox{70mm}{!}{\includegraphics{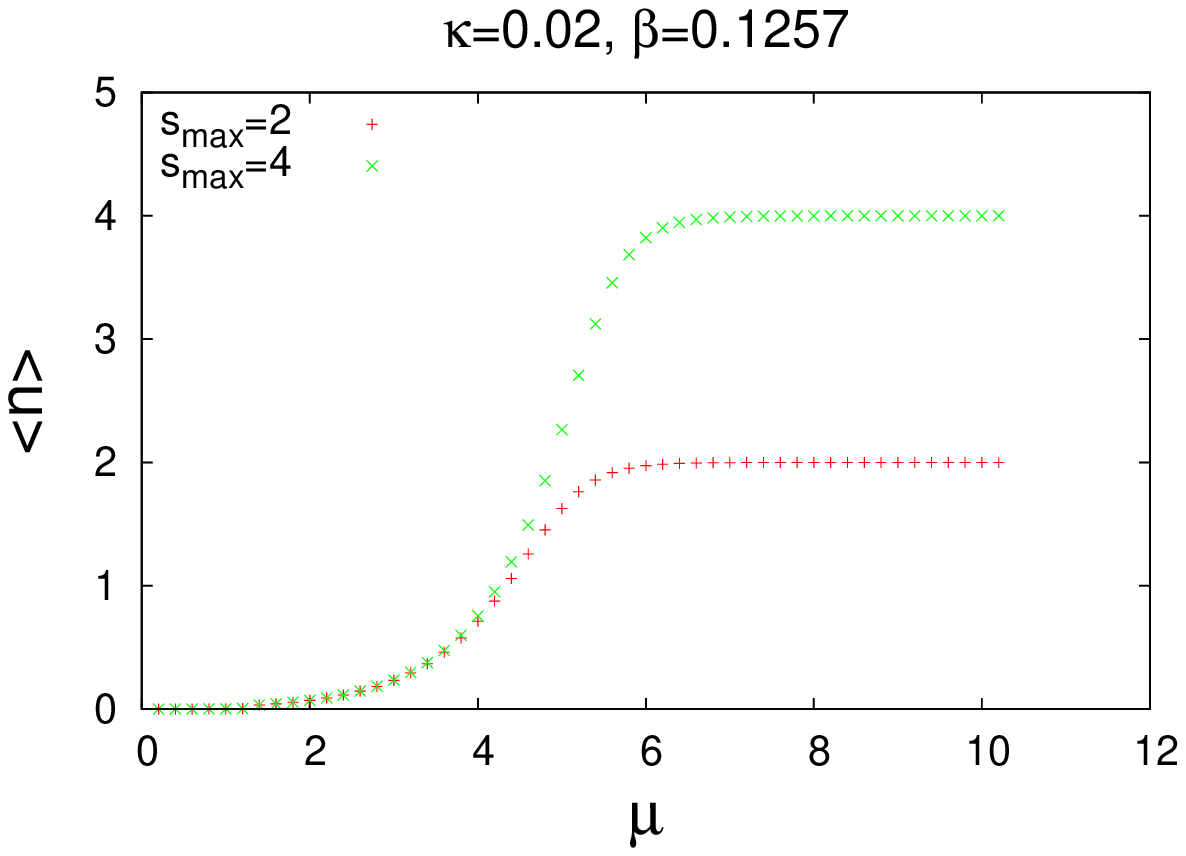}}
\label{n}
}
\caption{Observables $\langle \tr U \rangle, \langle \tr U^\dg \rangle$ and $\langle n \rangle$ vs.\ $\mu$ at fixed $\k=0.02$ and 
$\b=0.1257$, for two values of the cutoff $s_{max}$.  Note that these observables are independent of the baryon/site
cutoff $s_{max}$, until a little beyond $\m=4$, which is well past the value of $\m$ at the first order transition.}
\label{vsm}
\end{figure}

   The results, for $\mu \le 10$ and $s_{max}=2,4$, are shown in Fig.\ \ref{vsm}.  We see that $\langle \tr U \rangle$
and $\langle \tr U^\dg \rangle$ are comparable to one another and of $O(1)$ until $\langle n \rangle$ approaches
the cutoff in $s$.  Beyond that point, $\langle \tr U \rangle$ falls exponentially as $e^{-\m}$, and $\langle \tr U^\dg \rangle$
diverges as $e^\m$, exactly as in the U($N$) theory, and the results are no longer valid for the SU($N$) case.   When 
$\langle n \rangle$ saturates the cutoff then, in order to probe a larger range of $\m$, it is necessary to increase 
$s_{max}$.  For the purpose of determining the phase diagram, however, $s_{max}=4$, which can be interpreted as a limit of no more than four baryons per lattice site, appears to be more than sufficient.

\subsection{\label{corr}The leading correction to the mean field free energy}

    Going back to eq.\ \rf{DelF}, we have
\bea
e^{-\D F} = \langle e^J \rangle_{mf} =  \left\langle \prod_{x,k} e^{J_{x,k}} \right\rangle_{mf} \ .
\eea
The product is over all links, where
\bea
J_{x,k} &=&  \b  \Bigl\{ (\tr U_x^\dg -v)(\tr U_{x+\hk}-u) 
 + (\tr U_x -u)(\tr U^\dg_{x+\hk}-v) \Bigr\} \ ,
\eea
and the $\langle \rangle_{mf}$ notation denotes the expectation value with respect to the mean field action,
as in \rf{DelF}. The expansion of $\exp[J]$ generates products of terms such as $J_{l_1}J_{l_2}...J_{l_n}$,
where the $l_i$ denote links, some of which may be the same.
Because $\langle \tr[U] \rangle_{mf} = u$ and $\langle \tr[U^\dg] \rangle_{mf} = v$, it is clear that the expectation
values of such products are only non-zero if each endpoint of a link $l_i$ appearing in the product is also an endpoint
of at least one other link appearing in the product.  The simplest product whose expectation value is non-vanishing,
containing the minimum number of $J$ factors, is simply the product of $J_l J_l$ on the same link.  Therefore,
to leading order, we approximate
\bea
e^{-\D F} &=&  \left\langle \prod_{x,k} e^{J_{x,k}} \right\rangle_{mf}
\non \\
&\approx&  \prod_{x,k} \left\langle  e^{J_{x,k}} \right\rangle_{mf}
\non \\
&\approx&  \prod_{x,k}  (1+ \oh \langle J_{x,k}^2\rangle_{mf})  \ .
\eea
Now
\bea
\langle J^2_{x,k} \rangle_{mf} &=& \b^2  \left\langle (\tr U_x^\dg -v)^2 (\tr U_{x+\hk}-u)^2 \right. 
\non \\
& & \left. + (\tr U_x^\dg -v)(\tr U_x -u)(\tr U_{x+\hk}-u)(\tr U^\dg_{x+\hk}-v)  + h.c. \right\rangle_{mf} 
\non \\
&=& 2\b^2 \left[ \Bigl(\langle \tr U \tr U \rangle_{mf} - u^2 \Bigr)\Bigl(\langle \tr U^\dg \tr U^\dg \rangle_{mf} 
- v^2 \Bigr) 
  + \Bigl(\langle \tr U \tr U^\dg  \rangle_{mf} - uv \Bigr)^2 \right]
\non \\
&=& 2\b^2 \left[ \Bigl(\langle (e^{\m} \tr U)^2 \rangle_{mf} - u'^2 \Bigr)\Bigl(\langle (e^{-\m}\tr U^\dg)^2 
 \rangle_{mf} - v'^2 \Bigr) \right.
\non \\
& & \qquad  \left. + \Bigl(\langle (e^{\m} \tr U)(e^{-\m} \tr U^\dg)  \rangle_{mf} - u'v' \Bigr)^2 \right] \ ,
\eea
and we use
\bea
\Bigl\langle (e^\m \tr U)^m (e^{-\m}\tr U^\dg)^n  \Bigr\rangle_{mf} &=& {1\over G(A',B')} \left({\pa \over \pa A'}\right)^m 
      \left({\pa \over \pa B'}\right)^n G(A',B')  \ .
\eea
Putting all the pieces together, the free energy per unit volume is
\bea
f(A',B') &=& 2\b d u v - \tf(A',B') \ ,
\eea
where
\bea
\tf &=& \log G(A',B') + d\log\left[ 1 + \b^2 \left\{ \left( {1\over G} {\pa^2 G\over \pa A'^2} - u'^2\right) 
\left( {1\over G} {\pa^2 G\over \pa B'^2} - v'^2\right) \right. \right.
\non \\
& & \qquad \qquad \qquad \qquad \left. \left.+ \left( {1\over G} {\pa^2 G\over \pa A' \pa B'} - u' v'\right)^2
\right\} \right] \ ,
\eea
and $G(A',B')$ is as defined in \rf{G}.  Note that the terms inside the logarithm, which correct the leading mean field expression, depend on fluctuations around the mean field values.

    The variational parameters $A',B'$ are again derived by minimizing $f(A',B')$, which implies
\bea
           {B'-\k \over 2 \b d} - {\pa \over \pa A'} \tf &=& 0 
\non \\           
           {A'-\k \over 2 \b d} -  {\pa \over \pa B'} \tf &=& 0      \ ,    
\label{roots2}
\eea
whose roots may again be determined numerically.  It is also still true that $u=\langle \tr U \rangle, v= \langle \tr U^\dg \rangle$, which can be seen as follows:  Define
\bea
     \tZ \equiv e^{V \tf} = \int DU e^J \exp\left[ \sum_x (A \tr U_x + B \tr U^\dg_x) \right] \ .
\eea
Then it is clear that
\bea
 \langle \tr U \rangle = {1\over V} {\pa \over \pa A} \log \tZ  = {\pa \over \pa A} \tf \ .
\eea
Applying the first of eqs.\ \rf{roots2}, and the definitions \rf{rescale}, we arrive at $u=\langle \tr U \rangle$.
In the same way, we can show that $v= \langle \tr U^\dg \rangle$.  Thus the correspondence between the 
variational parameters $u,v$ and the observables  $\langle \tr U \rangle,  \langle \tr U^\dg \rangle$ is maintained exactly, in fact to all orders beyond the leading mean field expressions.

\begin{figure}[t!]
\centerline{\scalebox{0.80}{\includegraphics{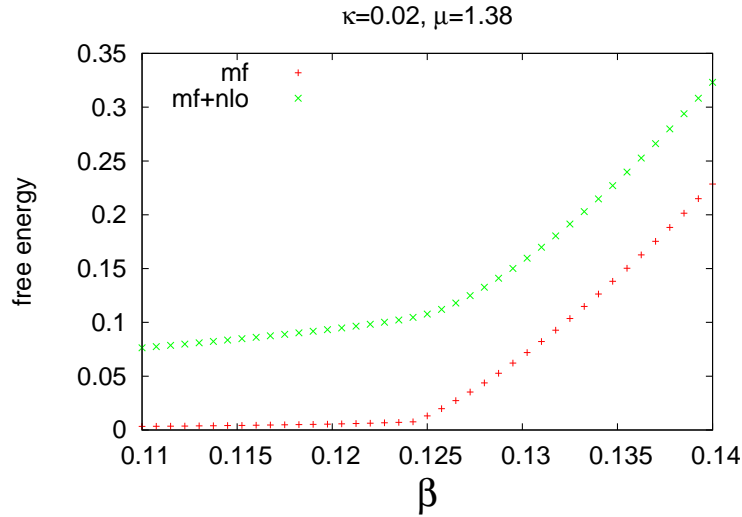}}}
\caption{$-F$ in lowest order mean field theory, and in mean field + next leading order.  Inclusion of the
next leading order can change a first-order transition to a crossover, as seen clearly in Fig.\ \ref{nlo_close}.  }
\label{nlo}
\end{figure}

\begin{figure}[ht]
\centering
\subfigure[~ mean field]{
\resizebox{70mm}{!}{\includegraphics{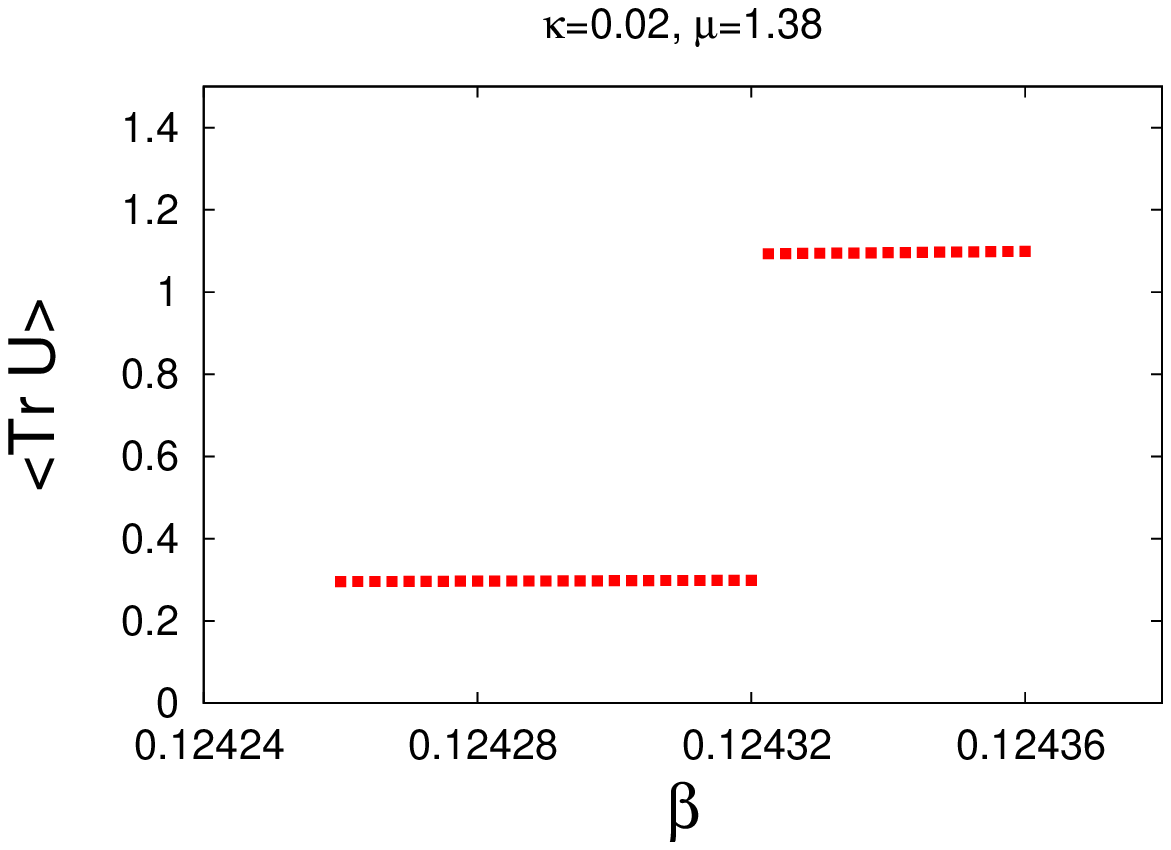}}
\label{uv}
}
\subfigure[~ mean field + NLO]{
\resizebox{70mm}{!}{\includegraphics{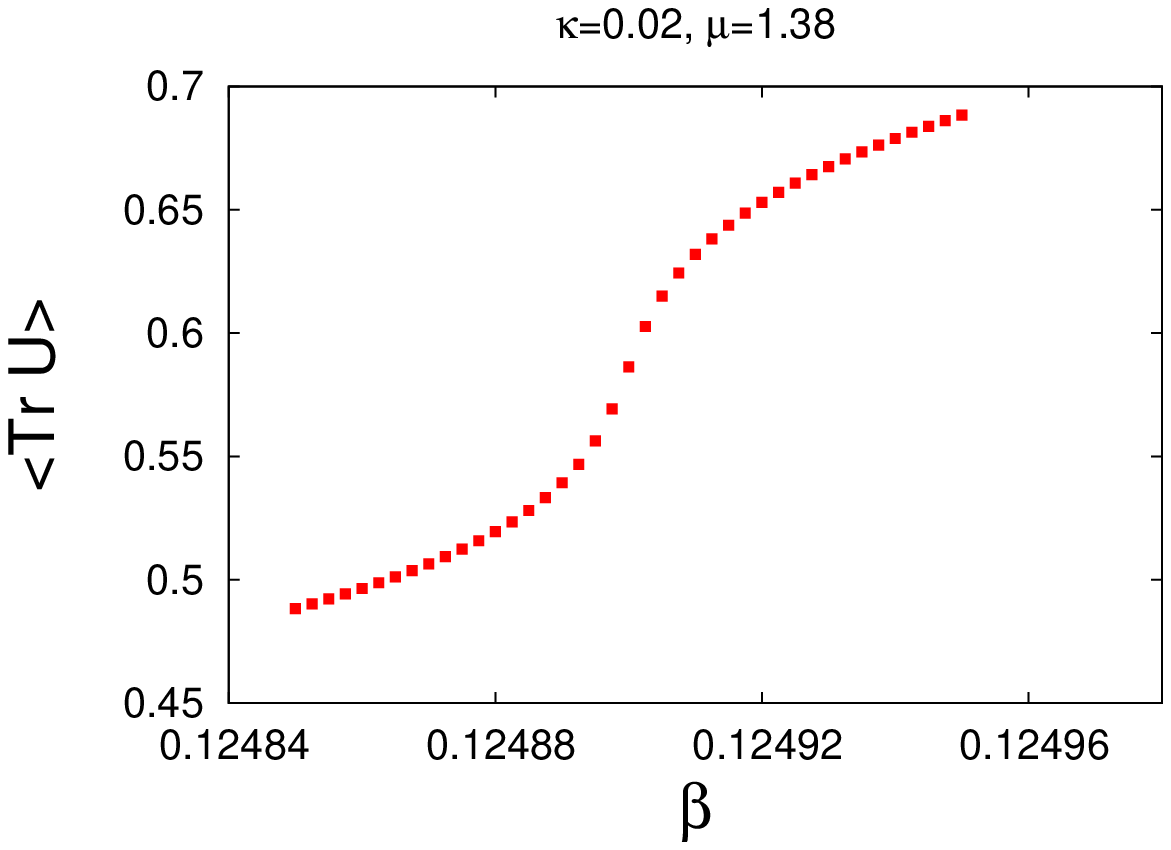}}
\label{free}
}
\caption{Closeup of $\langle \tr U \rangle$ in the transition region, at $\k=0.02$ and $\m=1.38$, showing the effect of inclusion of the leading correction to the mean field free energy.  With the inclusion of the leading correction, this
is the endpoint of the $\k=0.02$ line of transitions, down from the value $\m=1.67$, which is the endpoint without the first
correction.}
\label{nlo_close}
\end{figure}

   We can now study how inclusion of the leading correction will modify the phase diagram shown in 
Fig.\ \ref{phase}.   It turns out that the location of the phase transition points changes very little.   Generally,
at fixed $\k,~\m$, the value of $\b$ at the transition changes 
by less than one percent.  What does
change significantly are the endpoints of the first-order transition lines.  For example, at $\k=0.02$, the endpoint
of the transition line was at $\mu=1.67, \b=0.1213$.  Inclusion of the first correction brings the endpoint down to
$\mu=1.38, \b=0.1249$.  The free energy at lowest order (mf), and the free energy after inclusion of the first correction 
(mf+nlo) is shown in figure \ref{nlo}.  The free energy changes substantially, but the transition point hardly
at all (from $\b=0.1243$ to $\b=0.1249$).  However, at $\m=1.38$, the {\it order} of the transition changes, from first order in the leading mean-field approximation, to a sharp crossover when the first correction is included.  In figure 
\ref{nlo_close} we show a closeup of the $\langle \tr U \rangle$ in the near neighborhood of the transition in both cases.

   We also find that at $\k=0.04$, the endpoint of the line of first order transitions moves from $\m=0.87,~\b=0.1211$ to 
$\m=0.46,~\b=0.1246$.  At $\k=0.045$ the line of transitions shrinks to a point, at $\m=0,~\b=0.1245$.  Beyond 
$\k=0.45$, there are no transitions.

    So the first correction to mean field is taking us in the right direction, in the sense of bringing the endpoint of the
first order transition line to smaller values of $\m$.  Mercado and Gattringer \cite{Mercado:2012ue} find that the endpoints of the first-order transition lines are located at yet smaller values of $\m$.  It would be interesting to see if the next higher-order corrections generated by $\exp[J]$ would bring the endpoints still closer to the endpoints found in ref.\ \cite{Mercado:2012ue}.   We leave this exercise for a future study.

\section{\label{Proposal}Effective Spin Models and Full QCD}

    It seems to be easier to solve effective spin models at finite chemical potential, by a variety of methods, than to solve full QCD at finite chemical potential.  This means that if we knew the effective spin models corresponding to full QCD at relevant points in the plane of temperature and quark chemical potential, then by solving the effective models we could determine the QCD phase diagram. We know how to derive the effective spin model in the strong coupling and hopping parameter expansions; for $\mu=0$ this has been done in \cite{Green:1983sd,Ogilvie:1983ss}, and for $\m \ne 0$ in \cite{Fromm:2011qi}.  The problem is to go beyond these expansions, to weaker couplings and light quark masses, at $\mu \ne 0$.\footnote{For efforts at deriving the effective Polyakov line model in pure gauge theories, c.f.\ \cite{Wozar:2007tz} and references therein.}

    In principle, the effective Polyakov line model is derived from full QCD by integrating out the quark and gauge field variables, under the constraint that the Polyakov lines are fixed.  It is convenient to impose a
temporal gauge on the periodic lattice, in which all timelike links are set to the unit matrix except on a single time slice, $t=0$ say.  Then the effective theory, at chemical potential $\m=0$, is defined by integrating over all quark fields and link variables with the exception of the timelike links at $t=0$, i.e.
\bea
Z(\b,T,m_f) &=& \int DU_0(\bx,0) \int DU_k D\overline{\psi} D\psi ~ e^{S_{QCD}}
\non \\
                   &=& \int DU_0(\bx,0) ~ e^{S_{eff}[U_0,U_0^\dg]} \ ,
\label{S0}
\eea
where $\b$ is the gauge coupling, $T=1/N_t$ is the temperature in lattice units with $N_t$ the lattice extension in the time direction, and $m_f$ represents the set of quark masses.  Because temporal gauge has a residual symmetry
under time-independent gauge transformations, it follow that $S_{eff}$ is invariant under 
$U_0(\bx,0) \ra g(\bx) U_0(\bx,0) g^\dg(\bx)$, and therefore can depend on the timelike links only through their
eigenvalues.  This just means that $S_{eff}$ is a Polyakov line action of some kind.

   Let $S^{\m}_{QCD}$ denote the QCD action with a chemical potential, which can be obtained from $S_{QCD}$ by the following replacement of timelike links at $t=0$:
\bea
S^{\m}_{QCD} = S_{QCD}\Bigr[U_0(\bx,0) \ra e^{N_t \m} U_0(\bx,0), U^\dg_0(\bx,0) \ra e^{-N_t \m} 
U^\dg_0(\bx,0)\Bigl] \ .
\eea
The effective Polykov line action, at finite chemical potential is defined via
\bea
Z(\m,\b,T,m_f) &=& \int DU_0(\bx,0) \int DU_k D\overline{\psi} D\psi ~ e^{S^\m_{QCD}}
\non \\
                   &=& \int DU_0(\bx,0) ~ e^{S^\m_{eff}[U_0,U_0^\dg]} \ .
\label{Smu}
\eea

  The integration over $U_k,\overline{\psi},\psi$ can be carried out in a strong gauge-coupling  and hopping parameter
expansion, to obtain $S_{eff}$ and $S^\m_{eff}$.  It is not hard to see that each contribution to $S_{eff}$ in the strong coupling + hopping parameter expansion of \rf{S0} maps into a corresponding contribution to $S^\m_{eff}$, in the expansion of \rf{Smu}, by the replacement 
\bea
U_x \ra e^{N_t \m} U_x ~~~,~~~ U^\dg_x \ra e^{-N_t \m} U^\dg_x \ , 
\label{replace}
\eea
where we have identified $U_x\equiv U_0(\bx,0)$.
Since this mapping holds to all orders in the strong-coupling + hopping expansion, it is reasonable to assume that
 it holds in general, i.e.
 \bea
         S^\m_{eff}[U_x,U^\dg_x] = S_{eff}[U_x \ra e^{N_t \m} U_x, U^\dg_x \ra e^{-N_t \m} U^\dg_x] \ .
\label{mapping}
\eea

   Equation \rf{mapping} is a rather trivial, but potentially powerful identity.  It suggests that if, by some means, one could
obtain $S_{eff}$ at fixed $\{\b,m_f,T\}$ and chemical potential $\mu=0$, then one would immediately also have the effective action $S^\m_{eff}$ at the same set of parameters $\{\b,m_f,T\}$, but {\it any} chemical potential $\m$, by the replacement shown.  Unfortunately, there is some degree of ambiguity in $S_{eff}$ at $\mu=0$.  Suppose we have
some ansatz for $S_{eff}[U_x,U^\dg_x]$, depending on some small set of parameters, which we would like to fix by
comparing to the full theory at $\m=0$.   The problem is that whatever ansatz we make for $S_{eff}$, there is another form which is identical to that ansatz at $\m=0$, but differs under \rf{replace}.  In the case of SU(2), for example,
we could use the fact that $\tr U^\dg = \tr U$ to replace $\tr U_x^\dg$ by $\tr U_x$ (or vice versa) in $S_{eff}$.  Obviously, this will not affect anything at $\mu=0$, but will result in a different theory at non-zero $\mu$.  Similarly, in SU(3),
the identity
\bea
\tr U^\dg_x = \oh \Bigl[ (\tr U_x)^2 - \tr U_x^2 \Bigr]
\label{identity}
\eea
allows us to replace $\tr U^\dg_x$ everywhere in $S_{eff}$ by the right hand side of \rf{identity}, but this again produces
quite a different theory at $\mu \ne 0$ under the rule \rf{mapping}.  Of course a similar identity holds for $\tr U_x$, so we can convert the original $S_{eff}$ to another theory which may be symmetric in $U_x,~ U^\dg_x$, but which has quite a different extension to finite chemical potential.

   It may be possible to overcome this ambiguity, however.  Suppose we take the timelike link variables at $t=0$ to be
U(3), rather than SU(3) matrices.  Then the ambiguity due to \rf{identity} is no longer present, but the
effective spin theory still only depends on the eigenvalues of the U(3) matrices.  Then let us suppose that
we have some reasonable ansatz for $S_{eff}$ in a physically interesting range of parameters $\b,m_f,T$, e.g.  
\bea
         S_{eff} &=& \sum_{x,y} J(\bx-\by) \tr[U_x]\tr[U^\dg_y] + 
             \sum_{x,y} J'(\bx-\by) \Bigl(\tr[U_x]\tr[U_y]+\tr[U^\dg_x]\tr[U^\dg_y]\Bigr) + \sum_x V(U_x,U^\dg_x) \ ,
\non \\
\label{Seff2}
\eea
where $J(x),~J'(x)$ are parametrized by a few constants (such as nearest and next-nearest neighbor couplings), and
$V(U_x,U^\dg_x)$  can be limited to a few terms involving the characters of U(3).  In that case, the effective spin model is specified by a handful of constants $\{c_j\}$, which  of course depend on $\{\b,m_f,T\}$.

    Since there is no sign problem at $\m=0$ and $U_x = U_0(\bx,t=0) \in U(3)$, it should be possible to numerically simulate both the effective theory and the full theory.  Then one can imagine a number of strategies for obtaining the constants $\{c_j\}$.  One possibility is to simply calculate an appropriate set of observables in both theories (Polyakov lines in various representations and Polyakov line correlators), and fix the set of constants $\{c_j\}$ in $S_{eff}$ so that the two theories yield the same results.  Or perhaps some variant of the inverse Monte Carlo method could be applied \cite{Wozar:2007tz}.  A third procedure is inspired by a recent study of the Yang-Mills vacuum wavefunctional \cite{Greensite:2011pj}.  The idea is to select a finite set of $M$  timelike link configurations 
 \bea
 \{U^{(i)}_x = U^{(i)}_0(\bx,t=0) \in U(3), i=1,2,...,M\} \ ,
\eea 
where each member $U_0^{(i)}$ of the set specifies the timelike link variables at every spatial site $\bx$ and $t=0$.   Then the Monte Carlo simulation of the full theory proceeds in the usual way, except that on the $t=0$ timeslice, one member of the given set of timelike link configurations is selected by the Metropolis algorithm, and all timelike links on that timeslice are updated simultaneously.  Let $N_i$ be the number of times that the $i$-th configuration is selected by the algorithm, and $N_{tot} = \sum_i N_i$.  Then it is not hard to show that
\bea
             {\exp\Bigl[S_{eff}[U^{(i)}]\Bigr] \over \exp\Bigl[S_{eff}[U^{(j)}]\Bigr] } = \lim_{N_{tot}\ra\infty} {N_i \over N_j} \ .
\eea
Information derived from a number of such simulations, each using a different set of configuations at $t=0$, can in principle completely determine the $\{c_j\}$.  However, since the $\{N_i\}$ vary exponentially with $S_{eff}$, the variation of $S_{eff}$ within a given set must be kept relatively small, i.e.\ $\d S_{eff} \approx 5-7$, in order to ensure a reasonable acceptance rate for all members of the set.  For details of the algorithm, and its application to the vacuum wavefunctional of pure Yang-Mills theory, cf.\ \cite{Greensite:2011pj}.

    Once the set of constants $\{c_j\}$ is found, by whatever method, the effective theory at finite chemical potential, 
$S^\m_{eff}$, for any $\m$ but the same set $\{\b,m_f,T\}$, is given by the identity \rf{mapping}.   The final step
is simply to note that SU(3) $\subset$ U(3), so that the theory we want, $S_{eff}^\m$, is obtained by restricting the
$U_x$ matrices to the SU(3) subset.  Equivalently, since we can always express the U(3) matrices as \footnote{Allowing for the $Z_3$ subgroup of SU(3), the angle $\th_x$ can be restricted to the range $[0,2\pi/3)$.}
\bea 
U_x = \exp[i\th_x] U^{SU(3)}_x ~~~,~~~ U^\dg_x = \exp[-i\th_x] (U^{SU(3)}_x)^\dg
\eea
the conversion from $S_{eff}$ to $S^\m_{eff}$ is obtained by setting $\th_x=-i N_t \m$.

     With the effective Polyakov line model $S_{eff}^\m$ in hand, the theory can be solved by the mean field approach discussed above, or by other methods such as complex Langevin \cite{Karsch:1985cb,Bilic:1987fn,Aarts:2011zn}, the flux representation + worm algorithm \cite{Mercado:2012ue}, or reweighting \cite{Fromm:2011qi}.  To check that the method is working at $\m \ne 0$, one would compare full QCD with the effective spin model at, e.g., small or imaginary $\mu$, where the $\m$-dependence of the full theory can be obtained by other means.
     
      This approach can be expected to break down at sufficiently large $\mu$. At some point, terms in the potential involving high powers of $U_x$ and $U^\dg_x$, which might be negligible for computing observables at $\mu=0$ because they are multiplied by very small coefficients, could become important under the replacement  \rf{mapping}.   To what extent this effect will inhibit the study of the phase diagram remains to be seen.

    There is no doubt that determining the set of constants  $\{c_j\}$ in full QCD would be computationally demanding.
As a first step, it may be worth trying to extract the effective spin theory from gauge theories with scalar, rather than fermionic, matter fields.
    
\section{\label{Conclusions}Conclusions}

   The mean field expansion for effective spin models with a chemical potential turns out to have an interesting structure.
The constraint taking U($N$) to SU($N$) is responsible for the $\m$-dependence of the free energy, and this constraint introduces an infinite sum whose index, as it turns out, can be interpreted as the baryon number at each site.  The partition function can then be formally expressed in terms of a baryon fugacity expansion.

    If we ignore the distinction between first-order and crossover points, then even the lowest order mean field equations do a reasonably good job of accounting for phase structure.  The main error lies in the location of the endpoints of first-order transition lines, which occur, for fixed $\k$, at higher values of $\m$ than those determined by other methods.
The first correction to the mean field result moves those endpoints in the right direction, i.e.\ to lower values of $\m$.
It remains to be seen whether realistic results for the endpoints would be obtained from still higher orders in the mean field expansion.  

    We have also commented on the problem of deriving effective spin models from full QCD, and on certain subtleties associated with continuing those models from zero to finite chemical potential.  We have suggested that a method which was previously applied to study the Yang-Mills vacuum wavefunctional may be useful in this context, and hope to discuss this further at a later time.

\acknowledgments{We thank Poul Henrik Damgaard for useful discussions.  JG's research is supported in part by the
U.S.\ Department of Energy under Grant No.\ DE-FG03-92ER40711. The work of K.S.\ was supported by the Sapere
Aude program of the Danish Council for Independent Research.}

\bibliography{pline}

\end{document}